\documentclass[fleqn,12pt,twoside]{article}
\usepackage{espcrc1}

\usepackage{graphicx}
\usepackage[figuresright]{rotating}

\newcommand{\AmS}{{\protect\the\textfont2
  A\kern-.1667em\lower.5ex\hbox{M}\kern-.125emS}}

\newcommand{\be}{\begin{eqnarray}}

\newcommand{\ee}{\end{eqnarray}}

\def\simge{\mathrel{%
   \rlap{\raise 0.511ex \hbox{$>$}}{\lower 0.511ex \hbox{$\sim$}}}}
\def\simle{\mathrel{
   \rlap{\raise 0.511ex \hbox{$<$}}{\lower 0.511ex \hbox{$\sim$}}}}
\def\bigs{\mathrel{
   \rlap{\raise 0.531ex \hbox{$>$}}{\lower 0.531ex \hbox{$<$}}}}

\def\frac#1#2{{#1 \over #2}}

\def\half{\ifinner {\scriptstyle {1 \over 2}}
   \else {1 \over 2} \fi}

\title{The thermodynamics of the quark-gluon plasma:\\
Self-consistent resummations vs. lattice data}

\author{J.-P. Blaizot\address[SPHT]{Service de Physique Th\'eorique, CE Saclay,
        F-91191 Gif-sur-Yvette, France}, E. Iancu\addressmark[SPHT],
A. Rebhan\address{Institut f\"ur Theoretische Physik,
         Technische Universit\"at Wien,
         A-1040 Vienna, Austria}}
       
\begin{document}
\maketitle

\begin{abstract}
We discuss a recent approach for overcoming
the poor convergence of the perturbative expansion for the
thermodynamic potential of QCD. This approach is based on
self-consistent approximations which allow for a gauge-invariant 
and manifestly ultraviolet-finite resummation of the essential
physics of the hard thermal/dense loops. The results thus obtained
are in good agreement with available lattice data
down to temperatures of about twice the critical temperature.
Calculations for a plasma with 
finite quark density (i.e., with a non-zero chemical potential $\mu$)
are no more difficult than at $\mu=0$. 
\end{abstract}

\section{The failure of the conventional perturbation theory}

The most compelling theoretical evidence for the existence
of the quark-gluon plasma comes from lattice QCD which shows
a clear signal for a deconfinement phase transition \cite{Karsch}.
Above it, the lattice results slowly
approach the ideal-gas limit from below, with important deviations,
though, of about 15-20\%, up to temperatures $\sim 5T_c$,
and these are expected to remain noticeable
($\sim 10 \%$) even at temperatures as high as $10^3\, T_c\,$
\cite{BIR,Schroder}.

This suggests a picture of the high-temperature phase of QCD where
the interactions are more important than one would na\"{\i}vely 
expect on the basis of the asymptotic freedom alone, but where the
effects of these interactions are nevertheless small enough to be 
computable via weak-coupling techniques. 

The weak coupling expansion of 
the thermodynamic potential ${\cal F}$ (free energy) is present\-ly
known \cite{QCDP} to order 
$\alpha_s^{5/2}$, or $g^5$ ($\alpha_s\equiv g^2/4\pi$),
but it shows a disappointingly poor convergence except for coupling constants 
$\alpha_s \simle 0.05$ (which would correspond to 
 temperatures $\simge 10^5 T_c$).
Already the next-to-leading order correction of
${\cal O}(g^3)$ signals the inadequacy of the conventional
perturbation theory except for very small coupling,
because, in contrast to the leading-order terms, it leads to
a free energy in excess of the ideal-gas value.

This blatant inadequacy is somewhat surprising, since one
expects the 
free energy to be dominated by 
the {\it hard} thermal fluctuations with momenta 
$k\sim T$, for which perturbation theory should apply.
Indeed, at temperatures $T\simge T_c \sim 300$ MeV, the QCD 
coupling
is reasonably small, $\alpha_s\simle 0.3$,
when renormalized at the Matsubara scale $\bar\mu=2\pi T$.

But a closer inspection of the perturbative expansion reveals
that this is truly an expansion in powers of $g$ (rather than
$\alpha_s$), with $g\sim 1$ for all temperatures of interest,
and that the largest ``corrections'' 
are associated with {\it odd} powers of $g$.
The latter come from resummations which take into account the
phenomenon of Debye screening at the scale $gT$ \cite{PR}.
Thus, the large perturbative corrections are actually associated 
with {\it soft} degrees of freedom, with momenta of order $gT$,
and arise 
when the contribution of the
soft modes to the free energy is expanded in powers of $g$.
But this expansion is potentially troublesome, since, 
as we shall shortly recall, the soft modes are non-perturbative.

\section{The quasiparticle picture of the quark-gluon plasma}

The soft degrees of freedom are collective
excitations which would not even exist in the absence of interactions
\cite{PR}. To leading order in $g$, their dynamics
is described by an effective theory obtained by
integrating out the ``hard'' ($k\sim T$)
thermal fluctuations to one-loop order.
This generates a set of non-local  self-energy and vertex
amplitudes, known as ``hard thermal loops'' 
(HTL) \cite{BP}, which encompass screening effects
and non-trivial dispersion relations.
At momenta $k\simle gT$, the HTL's are {\it leading-order} effects
and must be resummed for consistency \cite{BP}.
Thus, when computing the contribution of the soft modes
to thermodynamical functions, these modes
should be treated as dressed {\it quasiparticles},
with properties described by the HTL effective theory.
This suggests a description of the thermodynamics of hot QCD 
in terms of weakly interacting ``hard'' and ``soft'' quasiparticles
(rather than the more strongly interacting elementary 
quanta).
This is also supported
by the success of phenomenological fits involving simple massive 
``quasiparticles'' in reproducing the lattice results \cite{Peshier}.

Quite generally, the  physical information on the quasiparticles
is contained in the spectral density $\rho(\omega, k)$
related to the corresponding propagator by:
\be\label{Dspec0}
D(\omega, k)&=&\int_{-\infty}^{\infty}\frac{{\rm d}k_0}{2\pi}
\,\frac{\rho(k_0, k)}{k_0-\omega}\,.\ee
For free massless excitations, $\rho_0(\omega, k)\propto
\delta(\omega^2-k^2)$. In the HTL approximation, 
the spectral densities are divided into a pole piece
at time-like momenta, and a continuum piece at space-like momenta. 
At  soft momenta $ k\simle gT$, all the pieces
of the HTL spectral functions appear to be equally important in our 
numerical calculations of thermodynamical quantities \cite{BIR}. 
At large momenta, the HTL spectral densities take the approximate form
\be\label{asymp}
\rho(\omega, k)\,\approx \,\delta(\omega^2-k^2-m^2_\infty)\qquad
{\rm for}\qquad k\sim T,
\ee
where $m^2_\infty\sim g^2T^2$ is the 
leading-order thermal mass (or ``asymptotic mass'') of
the {\it hard} excitations. Thus, quite remarkably,
the HTL approximation describes correctly the mass-shell
behaviour at both soft and hard momenta.  

In traditional perturbative calculations of the thermodynamics 
performed in imaginary time \cite{PR}, the HTL's play almost no 
role: only the Debye mass $m_D\sim gT$ needs to be resummed  
in the static ($\omega=0$) electric gluon propagator \cite{QCDP}.
Such a simple resummation retains
only one moment of the spectral function in eq.~(\ref{Dspec0}). 
Although this is enough
to obtain the leading order contribution,  $\propto g^3$,
of the soft modes to ${\cal F}$, it is clear that this 
procedure mistreats 
most of the physical content of the HTL's.

To overcome this limitation, two approaches have
been recently proposed to perform full resummations of the HTL 
self-energies in the calculation of thermodynamical functions
\cite{ABS,BIR}. In Refs. \cite{ABS}, this has been done
by merely replacing the free propagators by the
HTL-resummed ones in the expression of the free-energy of the ideal gas:
\be\label{ABS0}
{\cal F}_0\,=\,{1\over 2}\,
{\rm Tr} \log D_0^{-1} \,\longrightarrow \,
{\cal F}_{HTL}\,=\,{1\over 2}\,{\rm Tr} \log (D^{-1}_0+\Pi_{HTL})\,.\ee
In principle, this is just the first step in a systematic
procedure which consists in resumming the HTL's by adding and
subtracting them to the tree-level QCD Lagrangian.
This would be the extension to QCD of the so-called
``screened perturbation theory''\cite{KPP}, a method which, for
scalar field theories, has shown an improved convergence indeed,
in two- and three-loop calculations.
But in its one-loop approximation in eq.~(\ref{ABS0}), 
this method over-includes the leading-order interaction term
$\propto g^2$ (while correctly reproducing the order-$g^3$ 
contribution), and gives rise to new, ultimately
temperature-dependent, ultraviolet divergences
and associated additional renormalization scheme dependences.

Our approach on the other hand \cite{BIR} is based on self-consistent
approximations using the skeleton representation of the 
thermodynamic potential which takes care of overcounting problems
automatically, without the need for thermal counterterms.

\section{The (approximately) self-consistent entropy}

Specifically, we consider the 2-loop self-consistent 
(or ``$\Phi$-derivable'' \cite{Baym}) approximation
to the thermodynamic potential ${\cal F}$,
and focus on the {entropy}\footnote{More generally,
on the first derivatives of the thermodynamic potential, like
the entropy and --- for plasmas with non-zero chemical potential ---
also the quark density.}, which in this approximation takes
a simple, effectively one-loop, expression 
[with $N(k_0)=1/(\rm e^{\beta k_0}-1)$] :
\be\label{entropy}
{\cal S}\,=\,-\int\frac{{\rm d}^4k}{(2\pi)^4}\,\frac{\partial 
N}{\partial T}\, \left\{{\rm
Im}\ln D^{-1}\,-\,{\rm Im}\Pi[D]\,{\rm Re}D\right\}\,
\ee
but in terms of fully dressed propagators so that 
$\Pi[D]$ is the one-loop self-energy built out of the propagator $D$.
Thus, any explicit two-loop
contribution to the entropy has been absorbed into the spectral
properties of quasiparticles.
The price to be paid 
is that Dyson's equation 
$D^{-1}=D^{-1}_0+\Pi[D]$ becomes an integral equation, 
which is further complicated by UV problems and, 
in gauge theories, also by the issue of the gauge symmetry.

In spite of these complications, the expression (\ref{entropy}) has 
some obvious virtues: In addition to its simple quasiparticle
interpretation (as the entropy of a non-interacting gas of
quasiparticles with effective propagator $D$), it is
manifestly ultraviolet finite (the derivative of the statistical factor 
acting as an UV cut-off), and provides a non-perturbative
approximation to the thermodynamics which is perturbatively
correct up to, and including, ${\cal O}(g^3)$
(since the neglected 3-loop diagrams start contributing at
${\cal O}(g^4)$).

To cope with the problem that $\Phi$-derivable approximations
are not gauge invariant in general, we have proposed
gauge-independent but only approximately self-consistent 
dressed propagators as obtained
from (HTL) perturbation theory. Using these in eq.~(\ref{entropy})
gives a gauge-independent and UV-finite 
approximation for the entropy, which, while being
nonperturbative in the coupling, contains the correct leading-order
and next-to-leading order effects of the interactions.
Both arise from kinematical regimes where the HTL's are
justifiable approximations---at hard momenta, they involve
the HTL's only in the vicinity of the mass-shell, where the
HTL approximation is sound, cf. eq.~(\ref{asymp}).

\begin{figure}
\centerline{\includegraphics[bb=145 490 505 720,width=8cm]{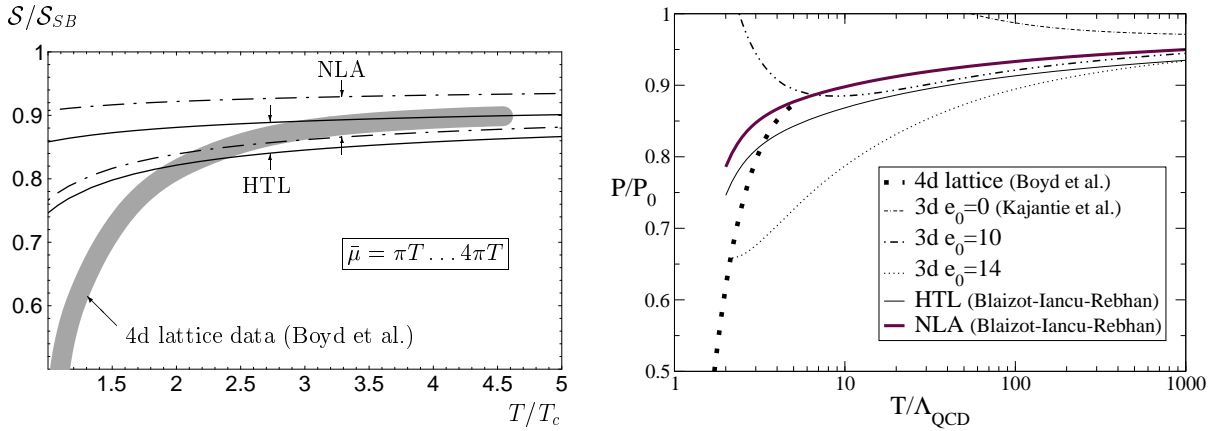}
\includegraphics[bb=10 75 710 525,width=8cm]{graceQM01.eps}\,}
         \caption{Entropy and pressure for pure-glue SU(3) Yang-Mills theory:
Approximately self-consistent
resummation \protect\cite{BIR} 
vs.
dimensional reduction \cite{Schroder}
and 4-d lattice data \cite{Karsch}.}
\label{S11}
\end{figure}

In the left half of Fig.~\ref{S11}, the results
for the entropy \cite{BIR} in a pure HTL approximation and in a
next-to-leading one (NLA) which includes soft corrections
to the hard asymptotic mass are compared with lattice data
for pure-glue SU(3) theory, showing good agreement for $T\simge
2.5 T_c\,$; in the right half, the resulting pressure\footnote{%
with renormalization scheme adjusted
so as to match that of Ref.~\protect\cite{Schroder}}
is compared in addition to the calculation of the pressure
in dimensional reduction of Ref. \cite{Schroder} for temperatures
up to $10^3 \Lambda_{\rm QCD}$. The fairly good agreement with the
best estimates of Ref.~\cite{Schroder} ($e_0=10$) seems to validate
our assumption of comparatively weak residual quasiparticle interactions.

Our method has been applied 
successfully also to plasmas with non-vanishing quark density 
(i.e., non-zero chemical potential) \cite{BIR},
for which lattice results are not yet available.


\begin{thebibliography}{9}

\bibitem{Karsch}
F. Karsch, hep-ph/0103314; 
G. Boyd {\it et al.}, Nucl. Phys. {\bf B469}, 419 (1996).

\bibitem{BIR} J.-P. Blaizot, E. Iancu and A. Rebhan, 
Phys. Rev. Lett. {\bf 83}, 2906 (1999); Phys. Lett. B {\bf 470},
 181 (1999); Phys. Rev. {\bf D63}, 065003 (2001).

\bibitem{Schroder}
K. Kajantie, M. Laine, K. Rummukainen, and Y. Schr\"oder, 
Phys. Rev. Lett. {\bf 86}, 10 (2001);
Y. Schr\"oder, proceedings of Quark Matter 2001.

\bibitem{QCDP}
P. Arnold, C. Zhai, 
Phys.\ Rev.\ D {\bf 50}, 7603 (1994),
{\it ibid.} {\bf 51}, 1906 (1995);
        C. Zhai, B. Kastening, {\it ibid.} {\bf 52}, 7232 (1995).
E. Braaten and A. Nieto, {\it ibid.} {\bf 53}, 3421 (1996).

\bibitem{PR} 
J.-P. Blaizot and E. Iancu, hep-ph/0101103 (and references therein).

\bibitem{BP} E. Braaten, R. D. Pisarski, Nucl. Phys. {\bf B337}, 569 (1990);
        J. Frenkel, J. C. Taylor, {\it ibid.} {\bf B334}, 199 (1990);
J.-P. Blaizot and E. Iancu, {\it ibid.} {\bf B417}, 608 (1994).

\bibitem{Peshier}
 P. L\'evai and U. Heinz, Phys. Rev. C {\bf 57}, 1879 (1998)
and references therein.

\bibitem{ABS} J. O. Andersen, E. Braaten, and M. Strickland, 
        Phys. Rev. Lett. {\bf 83}, 2139 (1999); 
        Phys. Rev. D {\bf 61}, 014017, 074016 (2000). 

\bibitem{KPP} F. Karsch, A. Patk\'os, and P. Petreczky, Phys. Lett. B {\bf401},
       69 (1997); 
J.~O. Andersen, E.~Braaten and M.~Strickland, hep-ph/0007159. 

\bibitem{Baym}
B. Vanderheyden and G. Baym, J. Stat. Phys. {\bf 93}, 843 (1998) and references therein.
\end{thebibliography}
\end{document}